# Effect of Pt doping on the critical temperature and upper critical field in YNi$_{2-x}$Pt$_x$B$_2$C (x=0-0.2)

Sourin Mukhopadhyay[a], Goutam Sheet[a*], A. K. Nigam[a], Pratap Raychaudhuri[a†] and H. Takeya[b]

[a]Department of Condensed Matter Physics and Materials Science, Tata Institute of Fundamental Research, Homi Bhabha Rd., Colaba, Mumbai 400005, India.

[b]National Institute of Materials Science, 3-13 Sakura, Tsukuba, Ibaraki 305-0003, Japan.

## Abstract

We investigate the evolution of superconducting properties by doping non-magnetic impurity in single crystals of YNi$_{2-x}$Pt$_x$B$_2$C (x=0-0.2). With increasing Pt doping the critical temperature ($T_c$) monotonically decreases from 15.85K and saturates to a value ~13K for x≥0.14. However, unlike conventional s-wave superconductors, the upper critical field ($H_{C2}$) along both crystallographic directions *a* and *c* decreases with increasing Pt doping. Specific heat measurements show that the density of states (N(E$_F$)) at the Fermi level ($E_F$) and the Debye temperatures ($\Theta_D$) in this series remains constant within the error bars of our measurement. We explain our results based on the increase in intraband scattering in the multiband superconductor YNi$_2$B$_2$C.

---

[*] Present affiliation: Department of Physics and Astronomy, Northwestern University, 2145 Sheridan Road, Evanston, IL 60208, USA

[†] Electronic mail: pratap@tifr.res.in

There has been an increase in interest multiband superconductors in recent years after the clear elucidation of multiband superconductivity[1] in MgB$_2$ arising from the π and σ bands. The simplest form of multiband superconductivity arises when electrons on different Fermi sheets in the same metal have different electron-phonon coupling strength leading to different sheets on the Fermi surface exhibiting different superconducting energy gaps (Δ). When the sample is in the clean limit, spectroscopic measurements on such a system reveal different superconducting energy gaps and different superconducting transition temperatures ($T_c$) on different Fermi sheets. Of particular interest in multiband superconductors is the evolution of the superconducting properties when one drives the system towards the dirty limit by substituting with non-magnetic impurities. In a conventional s-wave superconductor, non-magnetic disorder results in an increase in electronic scattering rate and decrease in the electronic mean free path (*l*). This results in a decrease in the coherence length and consequently an increase in the upper critical field[2] ($H_{c2}$) is observed. The $T_c$ on the other hand is not affected by non-magnetic impurities[3] unless the impurities result in a modification of the electronic or lattice properties, e.g. density of states at Fermi level ($N(E_F)$) or the Debye temperature ($\Theta_D$). In contrast, the situation in multiband superconductors is more complicated. For multiband superconductors in the clean limit, the band with strongest electron phonon coupling governs the bulk properties such as $T_c$ and $H_{c2}$. Substitution of non-magnetic impurities results in intraband scattering of electrons on individual Fermi sheets as well as interband scattering of electrons between different Fermi sheets. The former has an effect similar to conventional superconductor for individual bands, whereas the latter causes the bulk properties to be governed in the dirty limit by an average property of all the electrons instead of being governed by those with strongest electron-phonon coupling strength.

Submitted to Phys. Rev. B

Therefore, the evolution of $T_c$ and $H_{c2}$ for a multiband superconductor with substitution of non-magnetic impurities is governed by a complex interplay[4,5] of interband and intraband scattering. In particular, $H_{c2}$ need not necessarily increase with increase in electronic scattering. The evolution of $H_{c2}$ is a multiband superconductor with impurity is therefore a matter of considerable interest.

In this paper, we study the evolution $T_c$ and $H_{c2}$ in a series of Pt doped $YNi_2B_2C$ single crystals, e.g. $YNi_{2-x}Pt_xB_2C$ (x=0-0.2). The quaternary borocarbide superconductor[6] $YNi_2B_2C$ displays several unusual properties, namely, large anisotropy in the superconducting order parameter[7,8,9], a negative curvature[10] in $H_{c2}(T)$ close to $T_c$ and a square flux line lattice[11] at high magnetic fields. In a previous paper[12], through measurements of Δ using directional point contact spectroscopy (DPCS) as a function and magnetic field along different crystallographic directions, we have shown that these unusual properties can be understood from a multiband scenario, where large difference in the Fermi velocity on different Fermi sheets[13] gives rise to different electron-phonon coupling strength and different values of Δ. In this paper, we carry out a detailed measurements of $H_{c2}$, by applying magnetic field along the two crystallographic directions *a* and *c*. The central observations of this paper are the following: (i) With increasing *x*, both $T_c$ and $H_{c2}$ measured along the two crystallographic directions *a* and *c* decreases and saturates for x≥0.14; (ii) the anisotropy in $H_{c2}$ (for H||*a* and H||*c*) decreases monotonically with increasing *x*. Measurement of specific heat reveals that $N(E_F)$ and $\Theta_D$ does not change significantly within the error bars of our measurements. Our results elucidate the role of non-magnetic impurity in a multiband scenario where interband scattering dominates over intraband scattering.

Single crystals of $YNi_{2-x}Pt_xB_2C$ (x=0.02, 0.06, 0.1, 0.14 and 0.2) were grown by the traveling-solvent floating-zone method using an image furnace. X-ray powder diffraction using the crushed $YNi_{2-x}Pt_xB_2C$ single crystals were performed to determine the lattice parameters. X-ray profiles were analysed through Rietveld refinement using the FULLPROF program. A homemade high frequency (15KHz) planar coil a.c susceptibility setup was used to determine the critical temperatures ($T_c$) of these samples. The critical fields ($H_{C2}$) along two crystallographic directions *a* and *c* was determined using the same set up down to 2.2K and magnetic fields up to 8.5T. Specific heat measurements in the superconducting and normal state were carried out in a Quantum design Physical Properties Measurement System at zero field and at 9T respectively, over the temperature range 2-25K.

X-ray diffraction analysis reveals that pure $YNi_2B_2C$ has tetragonal lattice structure with lattice parameters *a*= 3.52Å and *c*= 10.54Å. With Pt doping the lattice parameters along both *a* and *c* axes increase monotonically and for x=0.2 the lattice constants along [100] and [001] become 3.54 Å and 10.62 Å respectively. The increase in the volume of the unit cell in the range x=0-0.2 is ~1.1%. Thus, there is a continuous incorporation of Pt on Ni sites without significant change in the atomic distances and the structural anisotropy[14]. Figure 1 shows the normalized real part of ac susceptibility ($\chi'$) as a function of temperature for all the samples. The $T_c$ is determined from the onset of superconducting transition, defined as 5% of the full signal change of the real part of a.c susceptibility ($\chi'(T)$). For the undoped $YNi_2B_2C$, $T_c$~15.85 K. With increased incorporation of Pt, $T_c$ decreases monotonically: $T_c$ falls sharply from 15.85 K (x=0) to 13.6 K (x=0.06) and then decreases gradually to 13K in the range x=0.06-0.2. Figure 2 shows $H_{c2//a}$ and $H_{c2//c}$ extracted from the field variation of $\chi'(H)$ for all the six samples as a function of temperature. The variation of χ'(H) with magnetic field at 2.2K is shown in the *insets* for both H||*a* and H||*c*. Almost all the χ'(H)-H graph shows a pronounced "peak effect"[15] (dip in χ'(H)) just before $H_{c2}$ arising from the order disorder transition of the vortex lattice. The $H_{c2}$ values at different temperatures are determined form the χ(T)-H data (inset Fig. 2(a-f)) using the same criterion as fixed to extract the $T_c$ values. To check the consistency of this procedure of extracting $H_{c2}(T)$, we have also compared the $H_{c2}(T)$ determined from isothermal M-H measurements[16] for the sample with x=0.2. The two curves are identical within the error bar of our measurement. We find a large anisotropy in the undoped YNi$_2$B$_2$C single crystal, with $\gamma_H = H_{C2||a}/H_{C2||c} > 1$. In pure YNi$_2$B$_2$C, $H_{c2//c} \sim 8.25$ T at 2.2K; for H||a, $H_{c2}$ is beyond our measurement limit of magnetic field at the same temperature. With increase in Pt, $H_{c2}(T)$ decreases in both the directions. The variation of $T_c$, $H_{c2}//a$ and $H_{c2}||c$ (at 2.2K) as a function of $x$ is shown in figure 3. The anisotropy decreases from $\gamma_H \approx 1.16$ for the sample with x=0.02, to $\gamma_H \approx 1$ for the sample with x=0.2 (*inset* Fig.3).

The rapid decrease in $H_{c2}$ in YNi$_2$B$_2$C upon substitution of Pt impurities is clearly not consistent with a conventional scenario. However, since the Fermi surface of YNi$_2$B$_2$C is very anisotropic we first compare the results with that of a single band superconductor with anisotropic Fermi surface. For a single band superconductor with anisotropic[2] Fermi surface in the clean limit,

$$H_{c2} \| c = \frac{\Phi_0}{2\pi (\xi_{0_{ab}})^2} = \frac{\Phi_0 \pi \Delta^2}{2\hbar^2 (v_{F_{ab}})^2} \quad (1)$$

$$H_{c2} \| a = \frac{\Phi_0}{2\pi \xi_{0_a} \xi_{0_c}} = \frac{\Phi_0 \pi \Delta^2}{2\hbar^2 v_{F_a} v_{F_c}}. \quad (2)$$

where $\Phi_0$ is the flux quantum and $v_{Fa}$ and $v_{Fc}$ are the Fermi velocities in the two directions (We assume $v_a=v_b=v_{ab}$ and $\xi_a=\xi_b=\xi_{ab}$ consistent with the tetragonal symmetry of the system). The anisotropy for such a superconductor given by $\gamma_H = v_{Fab}/v_{Fc}$ would gradually decrease with increase intraband scattering. However, the average value of the critical fields, $H_{c2}//a$ and $H_{c2}//c$ ($<H_{c2}>$) would show an increase due to reduction of electronic mean free path. In contrast, in YNi$_2$B$_2$C, in addition to the decrease in the individual values of $H_{c2//a}$ and $H_{c2//c}$, $<H_{c2}>$ decreases with increase in Pt doping to almost half its value in the clean limit.

To verify whether this evolution of $H_{c2}$ results from a change[17] in $N(E_F)$ or $\Theta_D$ upon substitution of Pt at the Ni site, we measured the specific heat ($C_p$) on the samples with x=0, 0.1 and 0.14. For all three samples measurements are carried out at H=0 and at H=9T, where the superconductivity is suppressed. Fitting the expression for the normal state specific heat[18], $C_n(T) = \gamma_n T + \beta T^3 + \alpha T^5$, (where $C_{electronic}=\gamma_n T$ and $C_{lattice}(T) = \beta T^3 + \alpha T^5$) with the $C_p$ measured at 9T the lattice contribution, $C_{lattice}(T)$ is evaluated. Since $C_{lattice}(T)$ is independent of magnetic field, the electronic specific heat ($C_{el}$) at H=0 is determined by subtracting the phonon contribution from the measured $C_p$ at H=0. Figure 4 shows the $C_{el}/T$ vs. T for three samples. It is clear that $C_{el}$ in the normal state does not change significantly showing that $N(E_F)$ is not affected by Pt doping. The extracted value of $\gamma_n$ and $\Theta_D$ for the three samples are (i) $\gamma_n$ =19±0.5mJ/mol K$^2$, $\Theta_D$=507±15 K for x=0; (ii) $\gamma_n$=20.1±0.5 mJ/mol K$^2$ and $\Theta_D$=522±15 K for x=0.1; (iii) $\gamma_n$=19.2±0.5 mJ/mol K$^2$ and $\Theta_D$=523±15 K for x=0.14. Though we could not measure the specific heat of the other composition due to the small mass of the crystals, within the error bars of our measurements, both $\gamma_n$ and $\Theta_D$ remain constant with Pt doping showing that the unusual variation of $H_{c2}$ does not result from modification of $N(E_F)$ and $\Theta_D$. Therefore, the variation, $H_{c2}$ in YNi$_{2-x}$Pt$_x$B$_2$C has to be analyzed beyond single band scenario.

To understand the variation of $H_{c2}$ with Pt doping we have to take into account the multiband nature of superconductivity in YNi$_2$B$_2$C. Spectroscopic measurements using DPCS on YNi$_2$B$_2$C single crystals[12] in the clean limit revealed that the presence of at least two groups of electrons on two different Fermi sheets, for which the superconducting energy gap and $T_c$ vary by a factor of 5-6. Comparison with band structure calculations[13] indicate that the 1$^{st}$ group of electrons are on a "square-pancake" Fermi sheet (SPFS), whereas the 2$^{nd}$ group of electrons are on a cylindrical Fermi sheet (CFS). In the clean limit, the bulk $T_c$ coincides with the 1$^{st}$ group of electrons with higher $T_c$. DPCS studies in the clean system indicates that $H_{c2}$ is also determined by the electrons on the "square-pancake" Fermi sheet[12,4] since the superconductivity on the other Fermi sheet is rapidly suppressed under the application of a magnetic field. Since the "square pancake" Fermi sheet is very anisotropic, for the clean system without significant interband scattering the $H_{c2}(0)$ will be given by equations (1) and (2) where $\Delta$ and $v_{Fab}$ and $v_{Fc}$ have to be replaced by the ones corresponding to the "square pancake" Fermi sheet. In such a system, addition of impurities would result[5] (i) in an increase in intraband scattering within each Fermi Sheet and (ii) an increase in interband scattering between the two Fermi sheets. The first effect will increase the bulk $H_{c2}$ governed by the coherence length of the electrons on the "square pancake" Fermi sheet, whereas the latter would decrease the superconducting energy gap (and $H_{c2}$) on the "square pancake" due to the influence of the cylindrical Fermi sheet. The rapid decrease of $<H_{c2}>$ with increase in Pt doping suggests that that the second effect dominates over the first one in the doping range of this study. This is expected since both the "square pancake"

Fermi sheet and the cylindrical Fermi sheet have large contribution from the Ni 3$d$ band. Pt doping at the Ni site is likely to increases the interband scattering between these two Fermi sheets. At the same time $\gamma_H$ decrease with increasing $x$ due to the decrease in anisotropy of individual bands.

Finally, we would like to note that the variation of $T_c$ with Pt doping can be understood from the same mechanism. With increase in interband scattering the $T_c$ will gradually decrease due to the influence of the 2$^{nd}$ group of electrons[19] with lower $T_c$, and will go towards a limiting value given by the weighted average of the $T_c$ of the two bands. The rapid decrease in $T_c$ at small values of $x$ and the subsequent leveling off for $x>0.14$ supports this scenario.

In summary, we have investigated the effect of impurities on $H_{c2}$ by studying a series of YNi$_{2-x}$Pt$_x$B$_2$C single crystals. We show that both $H_{c2}\|a$ and $H_{c2}\|c$ and the $T_c$ decreases with increasing x. Our results can be understood within a multiband scenario where the dominant contribution comes from the increase in interband scattering arising from Pt impurities. This study elucidates the role of interband scattering on the upper critical field in a multiband superconductor and reinforces the multiband nature of superconductivity in this material.

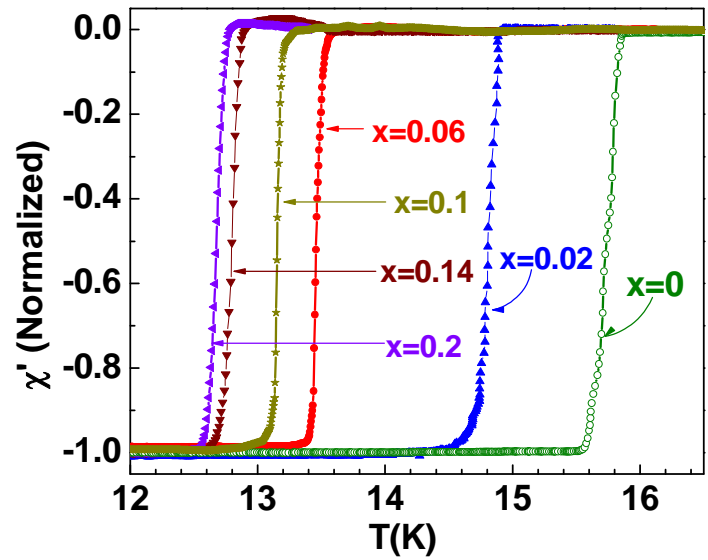

*Figure 1.* Normalized real part of a.c susceptibility ($\chi'$) as a function of temperature (T) in YNi$_{2-x}$Pt$_x$B$_2$C, for x=0-0.2. Solid lines are a guide to eye.

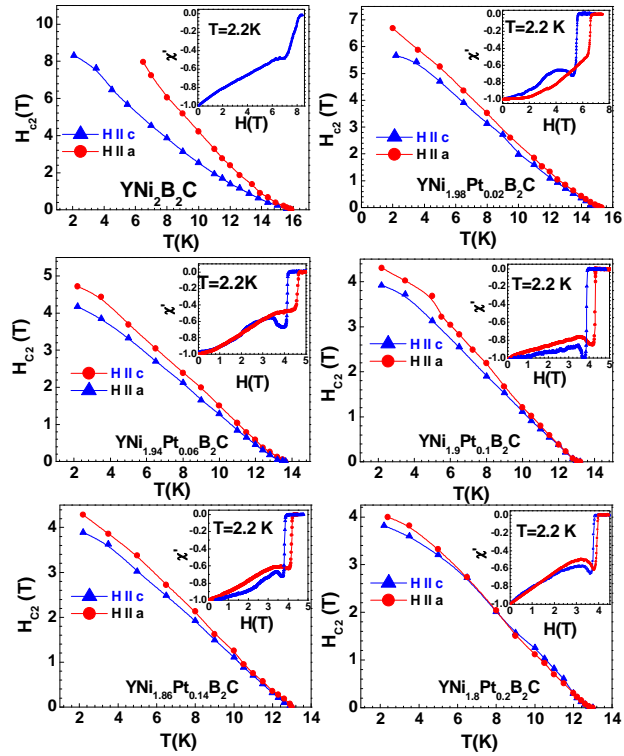

***Figure 2.*** Variation of $H_{c2}$ with temperature along H||a and H||c for $YNi_{2-x}Pt_xB_2C$, with x=0-0.2. Solid lines are a guide to eye. The *insets* show variation of normalized $\chi'(H)$ with magnetic field (H) at 2.2K for the same. All crystals show a pronounced "peak effect" close to $H_{c2}$.

Submitted to Phys. Rev. B

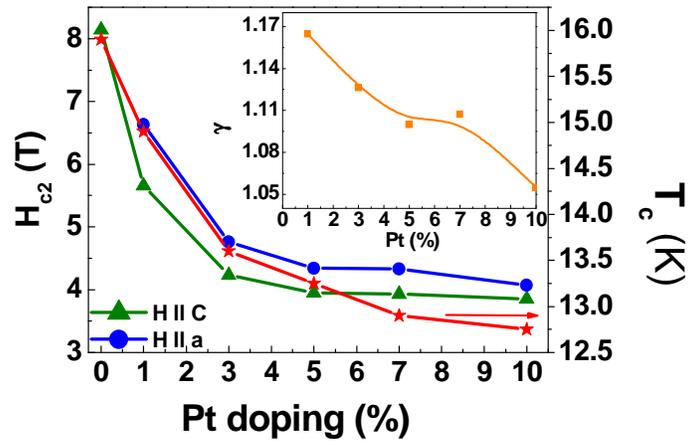

*Figure 3.* Variation of $H_{c2}$ with Pt doping in $YNi_{2-x}Pt_xB_2C$ in the range x=0-0.2, for both H||a (blue) and H||c (green). The red line shows the variation of $T_c$ with the same Pt doping. Solid lines are a guide to eye.

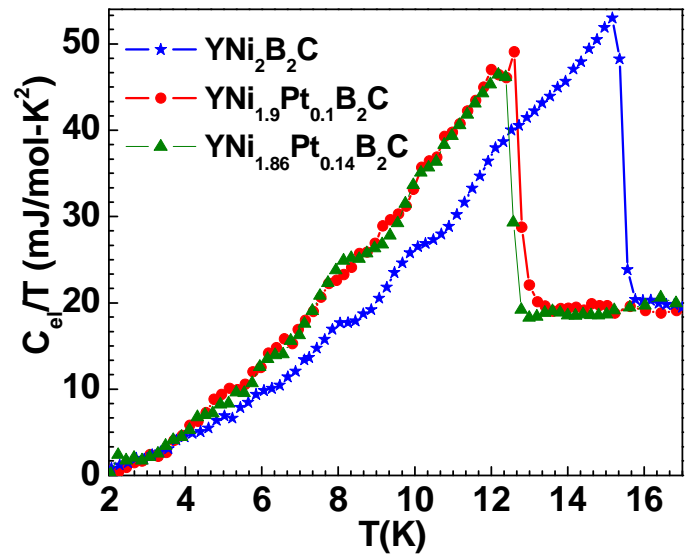

*Figure 4.* Temperature (T) dependence of electronic specific heat ($C_{el}$) plotted as $C_{el}/T$ vs. T for $YNi_{2-x}Pt_xB_2C$ with x= 0, 0.1 and 0.14. Solid lines are a guide to eye.

---

[19] H. Suhl, B. T. Matthias, and L. R. Walker, Phys. Rev. Lett. **3**, 552 (1959); E. J. Nicol and J. P. Carbotte, Phys. Rev. B **71**, 054501 (2005).